       \documentclass[12pt]{article}        

\usepackage{pslatex}
\usepackage{amsmath}
\usepackage{amssymb}
\usepackage{euscript}
\usepackage{epsfig}

\usepackage{cite}

\usepackage{alltt}

\usepackage{epic}
\usepackage{color}

\usepackage{enumerate}

\usepackage{subfigure}

\newif\ifpdf
\ifx\pdfoutput\undefined
    \pdffalse           
\else
    \pdftrue
\fi

\ifpdf
    \usepackage{pslatex}
\fi

\begin{document}                     

\allowdisplaybreaks

\begin{titlepage}

\begin{flushright}
 {\bf CERN-LCGAPP-2008-02}\\ 
 {\bf   IFJPAN-IV-2008-9 }
\end{flushright}
\vspace{1mm}
\begin{center}{\bf\Large  MC-TESTER v. 1.23: a universal tool}\end{center}
\begin{center}{\bf\Large   for comparisons of Monte Carlo predictions}\end{center}
\begin{center}{\bf\Large for particle decays in high energy physics $~^{\dag}$ }\end{center}
\vspace{3 mm}
\begin{center}
{\large \bf N. Davidson$^{a,d}$,}
{\large \bf P. Golonka$^{b}$,}
{\large \bf T. Przedzi\'nski$^{c}$,}
{\large \bf Z. W\c{a}s$^{d,e}$}\\
\vspace{3 mm}
{\em $^a$ University of Melbourne, Department of Physics \\ Australia.  }\\
{\em $^b$CERN, IT/CO-BE, CH-1211 Geneva 23, Switzerland.}\\
{\em $^c$ The Faculty of Physics, Astronomy and Applied Computer Science,\\
Jagellonian University, Reymonta 4, 30-059 Cracow, Poland.}\\
{\em $^d$Institute of Nuclear Physics,
         Radzikowsiego 152 , 31-342 Cracow, Poland.}\\
{\em $^e$CERN PH-TH, CH-1211 Geneva 23, Switzerland.}\\

\end{center}

\vspace{1mm}
\begin{abstract}
Theoretical predictions in high energy physics are routinely provided 
in the form of Monte Carlo generators.
Comparisons of predictions from different programs and/or different 
initialization set-ups are often necessary. 
  {\tt MC-TESTER} can be used
for such tests of decays of intermediate 
states (particles or resonances) in a semi-automated way.

 Since 2002 new functionalities
were introduced into the package. In particular, it now works with the
{\tt HepMC} event record, the  standard for C++ programs.
The complete set-up  for benchmarking the interfaces, such as
interface between $\tau$-lepton
production and decay, including QED bremsstrahlung effects is shown.
The  example is chosen to illustrate the new options introduced into the 
program.
From the technical perspective, our paper 
documents software updates and  supplements previous documentation.

As in the past,
our test consists of two steps. Distinct Monte Carlo
programs are run separately; events with decays of a chosen 
particle are searched, and  information is stored
 by {\tt MC-TESTER}. 
Then, at the analysis step,  information from a pair of  runs may
 be compared and represented in the form of tables and plots.

Updates introduced in the progam  up to version 1.24.3 are also documented.  
In particular, new configuration scripts or  script to combine
results from multitude of runs into single information file to be used
in analysis step are explained.

\end{abstract}
\begin{center}
{\it To be submitted to Computer Physics Communications, }
\end{center}

\vspace{1mm}
\begin{flushleft}
{\bf CERN-LCGAPP-2008-02 
}\\
 {\bf  IFJPAN-IV-2008-9  } 
\end{flushleft}

\vspace{-1mm}
\footnoterule
\noindent
{\footnotesize
\begin{itemize}
\item[${\dag}$]

This work is partially supported by  EU Marie Curie Research Training Network 
grant under the contract No. MRTN-CT-2006-0355505 and by  
Polish Government grant N202 06434 (2008-2010).

\end{itemize}
}

\end{titlepage}

\tableofcontents 

\newpage

\noindent{\bf \large PROGRAM UPDATE SUMMARY}
\vspace{10pt}

\noindent{\bf Title of the program:} \- {\tt MC-TESTER}, version 1.23
and version 1.24.3


\noindent{\bf Tested on  various platforms and operating systems:}\-
Linux SLC 4.6 and SLC 5, Fedora 8, Ubuntu  8.2 etc.

\noindent{\bf Programming languages used:}\- {\tt C++, FORTRAN77 }

\noindent{\bf Tested and compiled with:}\- gcc 3.4.6,  4.2.4 and 4.3.2 
with {\tt g77}/{\tt gfortran}

\noindent{\bf Size of the package:}\\
 23.4 MB  directory including example programs
 (6.5 MB compressed distribution archive), without
{\tt ROOT} libraries.

\noindent{\bf Additional disk space required:}  \\
Depends on the analyzed particle:
 14.4 MB  in the case of $\tau$ lepton decays (30 decay channels, 594 histograms,
 73-pages booklet). 


\noindent{\bf Nature of the physical problem:}\\
The decays of individual particles are  well defined
modules of a typical Monte Carlo program chain in high energy 
physics.
A fast, semi-automatic way of comparing results from different programs is 
often desirable for the development of  new programs, in order to check 
correctness of the installations or for discussion of  uncertainties.

\noindent{\bf Method of solution:}\\
A typical HEP Monte Carlo program stores the generated events in
event records such as {\tt HepMC}, {\tt HEPEVT} or  {\tt PYJETS}. {\tt MC-TESTER} scans, 
event by event, the contents of the record   and searches for the decays of 
the particle under study. The list of the found decay modes is successively 
incremented and histograms of  all invariant masses 
which can be calculated from the momenta of the particle decay products  
are defined and filled. The outputs from the two runs of distinct
programs can be later compared. A booklet of comparisons is 
created:  for every decay channel,
all histograms present in the two outputs are plotted and parameter
 quantifying shape difference 
is calculated. Its maximum over every decay channel is printed in the
 summary table. 

\noindent{\bf Restrictions on the complexity of the problem:}\-
Only first 200 decay channels that were found will initialize histograms and if the multiplicity of decay products in a given channel
was larger than 7, histograms will not be created for that channel as well. 

\noindent{\bf Typical running time:}\\
Varies substantially with the analyzed decay particle, but generally speed 
estimation of the old version remain valid. 
On a PC/Linux with 2.0 GHz processors 
{\tt MC-TESTER} increases the run time of  the $\tau$-lepton Monte Carlo
program   {\tt TAUOLA} by $4.0$ seconds for every $100~000$ analyzed events
(generation itself takes $26$ seconds).
The analysis step takes $13$ seconds; \LaTeX ~processing takes additionally $10$ seconds.
Generation step runs may be executed simultaneously on multi-processor machines.

\noindent{\bf New features:} {\tt HepMC} interface, use of lists in definition of 
histograms and decay channels, filters for decay products or secondary decays to be omitted, bug fixing, extended flexibility in representation of program output, installation configuration scripts, merging multiple output files from separate generations.


\noindent{\bf Accessibility: } \\
web page: {\tt http://mc-tester.web.cern.ch/MC-TESTER/} \\
e-mails:\hspace*{4.5mm} {\tt Piotr.Golonka@CERN.CH},
\\\hspace*{18.5mm} {\tt Zbigniew.Was@CERN.CH},
\\\hspace*{18.5mm} {\tt tomasz.przedzinski@uj.edu.pl},
\\\hspace*{18.5mm} {\tt Nadia.Davidson@CERN.CH}.

\noindent{\bf Reference to the program previous version: } \\
{P. Golonka, T. Pierzcha\l{}a, Z. W\c as, Comput. Phys. Commun., {\bf 157}(2004) 1}
\newpage

\section{Introduction}

In the phenomenology of high-energy physics, it is important to establish
uncertainties for theoretical predictions which are used in the interpretation 
of the experimental data is of high importance.
Theoretical predictions need 
to be presented in the form of Monte Carlo event generators; all detector 
effects can 
therefore be easily combined with the theoretical ones, using event
rejection or reweighting methods.
Whenever possible, theoretical predictions are separated into individual 
building blocks, which are later combined into complicated 
Monte Carlo generator systems for the complete predictions.

A good example of such a building block is  generator 
{\tt TAUOLA} \cite{Jadach:1993hs,Jezabek:1991qp,Jadach:1990mz}
for the simulation  of  $\tau$ lepton decay. In practical applications
such a program needs to be combined with other generators for the $\tau$ lepton
production,
{\tt TAUOLA universal interface} \cite{Pierzchala:2001gc,Golonka:2003xt}
can be then used. Additional complications arise due to  other effects 
such as final state bremsstrahlung,
 {\tt PHOTOS} Monte Carlo  
\cite{Barberio:1990ms,Barberio:1994qi,Golonka:2005pn,Golonka:2006tw}
can be then used. 

For the purpose of benchmarking our projects, we had to design and maintain
tests. Some of those tests
gradually evolved into the new version of  
{\tt MC-TESTER}~\cite{Golonka:2002rz} presented here.
The principle of these tests is rather simple. After generation of each event
by a given Monte Carlo system, the content is 
searched for the decay of the particle to be
studied. Once found, the appropriate data is collected 
and stored in the form of automatically created histograms and tables.

Originally the puropse of the  paper was to document {\tt MC-TESTER} 
 version  1.23. Recent improvements, for versions up to 1.24.3,
are described now as well. 
  For the properties of the program
 existing in even older versions, we address the reader to 
 ref.~\cite{Golonka:2002rz}. We will assume that the reader is
 familiar with that paper, otherwise technical aspects of the program 
explained here, may 
be difficult to follow.

For the convenience of readers interested in technical aspects of the update
we keep orders of the first chapters as in {\tt MC-TESTER}'s
first documentation \cite{Golonka:2002rz}:
\begin{itemize}
    \item Section \ref{sec:installation} explains updates introduced
to the first (generation) step of the
program.  Comments on the  installation procedure are also given here.
    In Section \ref{sec:analysis}  modifications introduced
in the analysis, the second step of
{\tt MC-TESTER} operation, are explained. 
    \item Section \ref{sec:package-organization} is devoted to the description
of the package update. In particular, directory organization and technical information on its use;
further details and explanation of input parameters may be found in
the appendix \ref{appendix.Setup-parameters}.
     Sections \ref{other-gen:C++}, \ref{sec:HepMCaid} and Appendix
\ref{appendix.C++} are devoted to the extension of {\tt MC-TESTER} to {\tt HepMC} and C++
applications.
    \item Section \ref{sec:Lists} is devoted to the use of lists in the algorithms
responsible for defining histograms and decay channels.
    \item New options and examples of how to obtain refined numerical results 
with {\tt MC-TESTER} are
 explained in Section \ref{sec:Benchmarks}.

    \item Section \ref{sec:outlook} closes the documentation with a discussion
of the package limitations and possible future extensions.
    \item Changes in configuration scripts 
and script to merge several {\tt MC-TESTER} output files 
introduced respectively for version 1.24.2 and 1.24.3
are documented in Appendix \ref{appendix.B}.

\end{itemize}
\section{Installation and generation step (update for ref.~\cite{Golonka:2002rz}) }
\label{sec:installation}

{\tt MC-TESTER} is distributed in a form of an archive containing source files.
Currently only the Linux and Mac OS\footnote{For this case LCG configuration 
scripts explained in Appendix \ref{appendix.B} have to be used.} operating systems are supported: other systems may be
supported in the future if sufficient interest is found. We have checked {\tt MC-TESTER}
on various platforms such as Scientific Linux SLC 4.6 or  Ubuntu 7.10, 8.04 .

In order to run {\tt MC-TESTER} the following software is needed:
\begin{itemize}
    \item {\tt gcc}\footnote{{\tt MC-TESTER} has been tested in particular with gcc  3.2, 4.03, 4.1.2 and 4.3.2}
      compiler suite with {\tt g++} and {\tt g77}/{\tt gfortran} installed.
    \item {\tt ROOT} package properly installed and set up (please refer to 
      \cite{root-install-www}  or {\tt ROOT\_INSTALL} file in {\tt doc/} 
      subdirectory for details),
    \item \LaTeX\  package,
    \item Version 1.23 requires  that environmental variables {\tt ROOTSYS} 
and {\tt LD\_LIBRARY\_PATH}, (for some applications also {\tt HEPMCLOCATION}) 
are properly set. It is optional to set those variables manually for 
version 1.24.3. The new installation procedure is explained in 
Appendix \ref{appendix.B}.
\end{itemize}
One  compiles {\tt MC-TESTER} libraries using
the {\tt make} command to be executed in its main directory\footnote{
 The {\tt MC-TESTER} version 1.23 is set to be compiled using g77 compiler. 
For {\tt gfortran} compiler, logical link {\tt make.inc} has to be pointed
(in directory   {\tt platform}) to   {\tt make-gfortran41.inc}. The 1.24.3 version does not require manual compiler setup.
}. If completed successfully, 
the user is instructed on how to proceed with
the example tests\footnote{Also, how to  prepare additional libraries to be loaded 
into user programs.}. Examples for {\tt MC-TESTER} use, based on 
the $\tau$ decay generators {\tt TAUOLA} and {\tt PYTHIA}
are distributed together with the package; they reside in the {\tt examples-F77/}
subdirectory.

{\tt MC-TESTER} distribution is a complete, ready-to-use testing environment, with subdirectories
dedicated to generation and analysis steps (see  Section \ref{dir-struct} for details),
and run-time parameters controlled by simple configuration files
({\tt SETUP.C} - see Section \ref{SETUP.C} and the Appendix \ref{appendix.A}). 
 The {\tt SETUP.C} file needs to 
be put in the directory from which the generation program is being executed 
(usually it is the same directory in which
the binary executable file exist). Examples of {\tt SETUP.C} files are already present
in example generation directories: they are used to set some parameters and 
also to note the name and details of
the generator being run.

The output data file is usually put in the directory in which the generation program was
executed. The name of the file and the path may however be changed using {\tt SETUP.C}.

The issue of using {\tt MC-TESTER}  with ``any'' Monte Carlo generators is addressed
in Section \ref{other-generators}. We want to stress, that it is relatively easy to use {\tt MC-TESTER}
with  a Monte Carlo event generator: it is sufficient to 
link the {\tt MC-TESTER} libraries, the {\tt  ROOT} libraries and to insert 
three subroutine calls into 
the user's code: for {\tt MC-TESTER} initialization, finalization and analysis.

For the users interested in trying only the analysis part of {\tt MC-TESTER} 
(Section \ref{sec:analysis}), and to avoid a lengthy generation phase,
ready-to-use data files are provided in the directory \\
{\tt examples-F77/pre-generated/}. There, the {\tt MC-TESTER}'s 
{\tt mc-tester.root} files
(produced by long runs with {\tt TAUOLA} and {\tt PYTHIA}), are stored. To copy the files to 
the directories of the analysis step, 
the command {\tt make move} can be used. In principle these files can be used 
as a reference for benchmark purposes too.

\subsection{Examples of C++ generation step}

Demonstrations of {\tt MC-TESTER}'s usage with C++ generation programs 
can be found
in the sub-directory {\tt example-C++/}. This includes an example of $\tau$ decay analysis
for {\tt PYTHIA 8.1} \cite{Sjostrand:2007gs} (using C++ and the {\tt HepMC} 
\cite{Dobbs:2001ck} standard) and an example of B meson decay
for {\tt EvtGenLHC} \cite{EvtGenLHC} (using C++ and the HEPEVT standard).
The examples are chosen, to demonstrate  the program use
for  packages of  widespread popularity.

\subsubsection{Tau decays from {\tt PYTHIA 8.1}}
An example for the {\tt PYTHIA 8.1} event generator is given in  directory {\tt examples-C++/pythia/}. 
10,000 $e^+ e^- \rightarrow Z^0 \rightarrow \tau^+ \tau^- $ events are generated and the decay of taus 
are analyzed by {\tt MC-TESTER}. The output, {\tt mc-tester.root}, contains results of 
the processed events and includes a number of histograms which can be compared to 
similar output from other Monte-Carlo generators. 
Configuration of the tool is done via the {\tt examples-C++/pythia/SETUP.C} file.

To run the example, the packages {\tt PYTHIA 8.1} and {\tt HepMC} version 2 have to be 
installed\footnote{This example has been tested with {\tt PYTHIA} version 8.100 and 
{\tt HepMC} versions 2.01.08\;{}-\;{}2.05.00. We assume that the reader is familiar with these packages and their documentation.}. 
{\tt PYTHIA 8.1} has to be compiled with {\tt HepMC} and the {\tt PYTHIA} library {\tt libhepmcinterface} 
must exists.
To run the example with {\tt MC-TESTER} version 1.23:
\begin{itemize}
  \item Set environment variable {\tt HEPMCLOCATION} to the base of {\tt HepMC}'s {\tt include/} and {\tt lib/} directories
  \item Set {\tt PYTHIA\_INSTALL\_LOCATION} to the base of {\tt PYTHIA 8}'s {\tt include/} and {\tt lib/} directories
  \item The {\tt PYTHIA8DATA} should point to directory containing {\tt PYTHIA} xml documents. Generally these can be 
    found in {\tt \$(PYTHIA\_INSTALL\_LOCATION)/xmldoc}.
  \item Compile the interface library by executing {\tt make libHepMCEvent} in the base directory of {\tt MC-TESTER}.
  \item Compile the example by executing {\tt make} in {\tt examples-C++/pythia} subdirectory.
\end{itemize}
In case of {\tt MC-TESTER} version 1.24.3:
\begin{itemize}
  \item Provide the location of {\tt HepMC} during configuration step
  \item Compile the {\tt MC-TESTER} libraries.
  \item {\tt Configure} example in {\tt examples-C++/pythia}; provide path to {\tt PYTHIA 8.1}.
  \item Compile  with {\tt make} command.
\end{itemize}
For details of {\tt HepMC} and {\tt PYTHIA 8.1}  see~\cite{Dobbs:2001ck,Sjostrand:2007gs}. For details regarding the configuration procedures see Appendix \ref{appendix.B}.

In order to run the example enter  {\tt examples-C++/pythia} directory and execute:\\
\\
{\tt ./pythiatest.exe}\\
{\tt make move1} (or {\tt make move2})\\

The final step moves the output file, {\tt mc-tester.root}, to the directory {\tt /analyze/prod1} 
({\tt /analyze/prod2}) ready for the 
analysis step. A second output should be produced using (preferably) different Monte Carlo generator (for example from 
{\tt PYTHIA 6.4} generation) and moved to {\tt /analyze/prod2}. 

\subsubsection{B decays from {\tt EvtGenLHC}}
An example of the analysis of 10,000 $B^+$ decays can be found in the sub-directory \\
{\tt examples-C++/evtgenlhc/}. This example requires {\tt MC-TESTER} to be linked with the libraries 
of {\tt EvtGenLHC}, {\tt PYTHIA}, {\tt PHOTOS}, {\tt CLHEP} and {\tt StdHep}
\footnote{This example has been tested with {\tt EvtGenLHC} version 5.15, {\tt PYTHIA} version 6.227.2, {\tt PHOTOS} version 215.5, {\tt CLHEP} version 1.9.3.1 and {\tt CERNLIBS} 2006 on afs at CERN. See web pages of 
LHC Computing Grid Project Generator Services Subproject http://lcgapp.cern.ch/project/simu/generator/, 
LCG Savannah https://savannah.cern.ch/projects/clhep/ and web page 
for Scientific Linux Installation at main CERN cluster
http://plus.web.cern.ch/plus/SLC4.html}. 
The path to each must be set with the following environmental variables: 
\begin{itemize}
\item {\tt EVTGEN\_INSTALL\_LOCATION}
\item {\tt PYTHIA6\_INSTALL\_LOCATION}
\item {\tt PHOTOS\_INSTALL\_LOCATION}
\item {\tt CLHEP\_INSTALL\_LOCATION}
\item {\tt CERNLIBS\_INSTALL\_LOCATION}
\end{itemize}

The {\tt EvtGenLHC} example should be run in the {\tt examples-C++/evtgenlhc/} directory 
with the following commands:\\\\
{\tt make}\\
{\tt ./evtgen\_test.exe}\\
{\tt make move1} (or {\tt make move2)}\\

The $B^+$ meson decays need to be generated with 
the help  of another  Monte Carlo generator to be  compared to the
results of {\tt EvtGenLHC}. For that purpose our 
example  \\
{\tt examples-C++/pythia/pythia\_test\_B.cc} can be used.
The resulting booklet is particularly large, confirming technical robustness
of {\tt MC-TESTER}.

\section{Analysis (update for ref.~\cite{Golonka:2002rz}) }
\label{sec:analysis}
Data files  {\tt mc-tester.root}, referred to in
 the previous section are used to produce 
a booklet - a final results of the  {\tt MC-TESTER} executions.
For this purpose the directory {\tt analyze/} is prepared. No significant changes 
in the analysis step of {\tt MC-TESTER} were introduced since its first public release.
The original documentation, given in~\cite{Golonka:2002rz} is to a large degree up to date.
In the following subsections we present minor changes which were nonetheless introduced. 

\subsection{Running the analysis step from an external directory }
In some cases, for example when our program  is installed centrally 
by an administrator,
 users may not have write permission for the {\tt analyze/} directory
of\; {\tt MC-TESTER}.
Therefore the analysis step will need to be executed from a directory outside {\tt MC-TESTER.}
The bash script \\ {\tt analyze/compare.sh} demonstrates how this can be done.
It should be copied to the working directory, and then edited. The following 
need to be set:
\begin{itemize}
  \item {\tt FILE1}: The name of the output file from the first generator (eg. {\tt mc-tester.root}). The path should be
    given relative to the current working directory.
  \item {\tt FILE2}: The name of the output file from the second generator.
  \item {\tt MCTESTER\_DIR}: The path to {\tt MC-TESTER}
\end{itemize}
{\tt MC-TESTER} can be configured by a {\tt SETUP.C} file in the working directory (See Section \ref{SETUP.C}).
If this file is not present, the default settings from {\tt analyze/SETUP.C} are used.

Execution of {\tt compare.sh} will produce an analysis booklet 
in {\tt pdf} format called {\tt tester.pdf}. Other files and directories produced 
during the analysis step, for example {\tt ROOT} histograms, can also be found in the 
user's current working directory.

\subsection{New options in graphical representation of histograms }
\label{sec:Graphical}


Graphical representation of the plots, as present already  in the original 
version of {\tt MC-TESTER} is adequate for tests, if agreement 
between the two compared Monte Carlo samples confirms.  
Generally it is however not the case. The appropriate graphical representation of tests 
can be then helpful to understand the origin or nature of differences.
The present version of {\tt MC-TESTER} introduces a few additional options 
that make histograms more readable in specific cases. 

\begin{figure}[!ht]
\setlength{\unitlength}{0.1mm}
\begin{picture}(1600,800)
\put( 375,750){\makebox(0,0)[b]{\large }}
\put(1225,750){\makebox(0,0)[b]{\large }}
\put(5,100){\makebox(0,0)[lb]{\epsfig{file=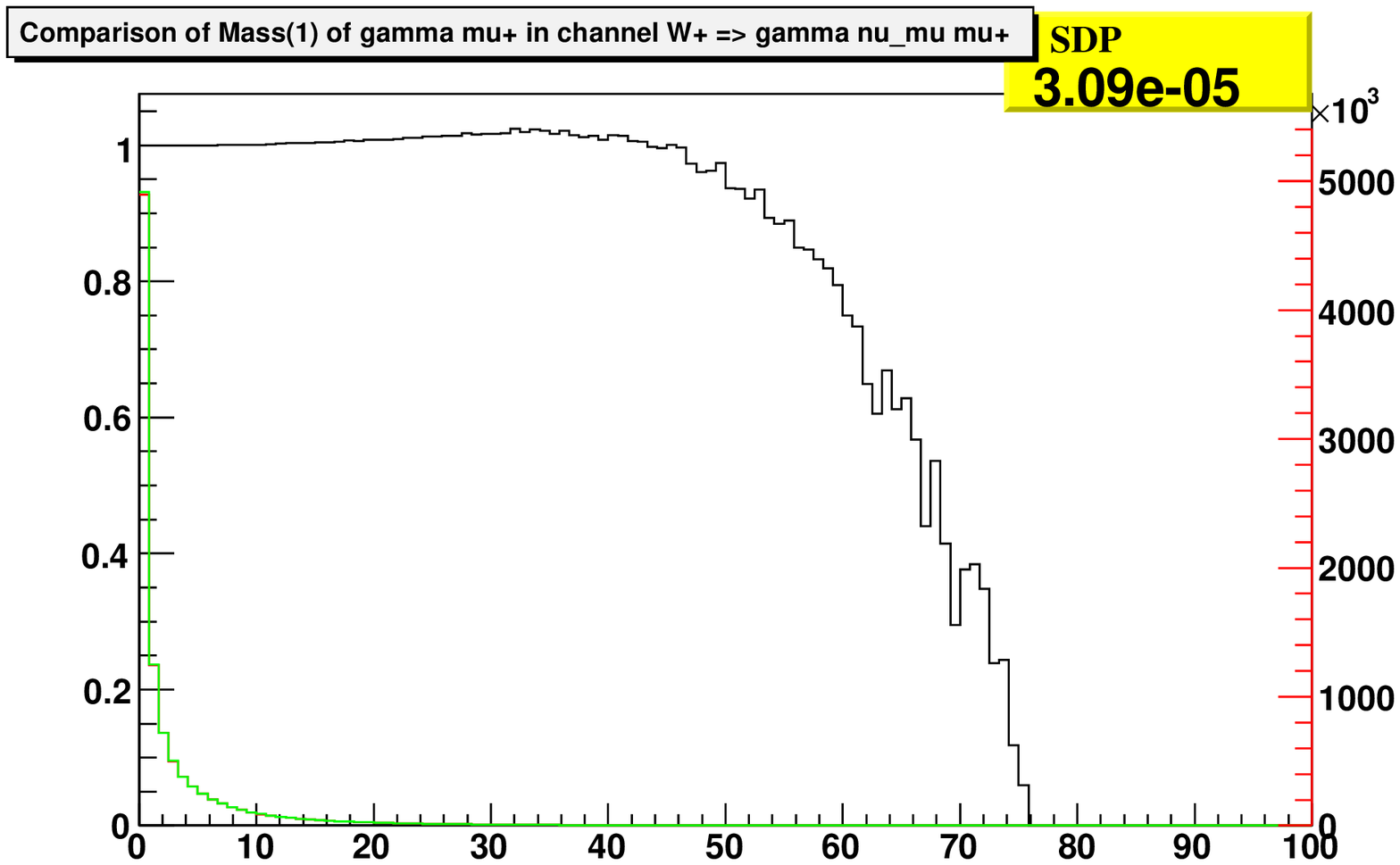,width=75mm,height=60mm}}
}
\put(55,75){\makebox(0,0)[lb]{a}
}
\put(905,100){\makebox(0,0)[lb]{\epsfig{file=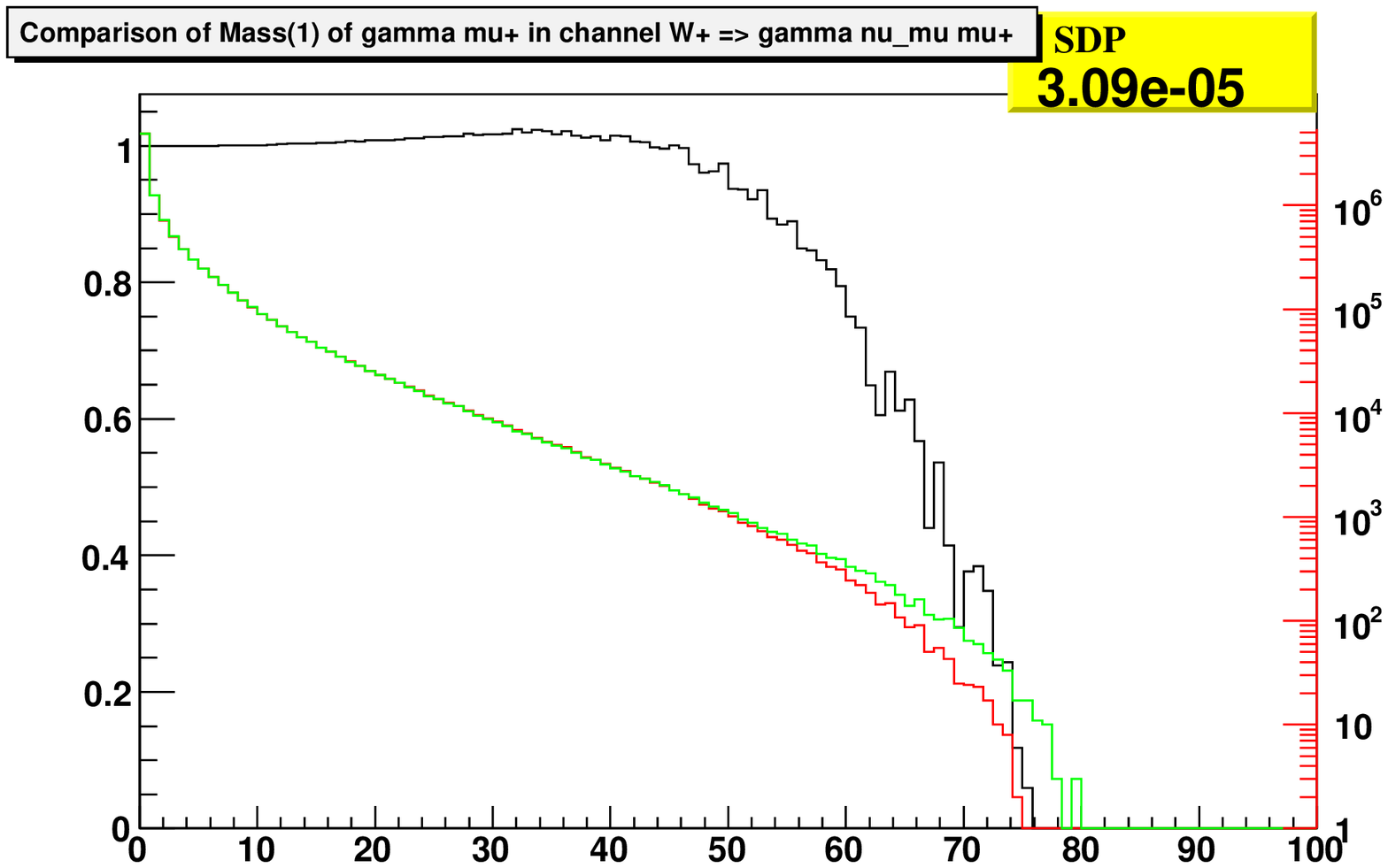,width=75mm,height=60mm}}
}
\put(955,75){\makebox(0,0)[lb]{b}}
\end{picture}
\caption 
{\it An example of a plot used for testing the full matrix element kernel of the
$W^+ \to \mu^+ \nu_\mu \gamma$ channel. The invariant mass distribution of 
the $\mu^+ \gamma$ pair is peaked on the left side, making the distributions unreadable.
In particular, the difference between the red and green lines can not be seen
at all. 
Only the ratio of the compared distribution is of some use. A theoretically unprepared
reader may get the impression that differences are large. 
 If logarithmic scale is used,
see fig.~$b$, invariant mass distributions are visible over the
whole spectrum and the difference is clearly localized in a region containing
less than $10^{-3}$of the whole sample. Calculation of SDP is not affected
by the rescaling of the histograms. }
\label{logfalse}
\end{figure}

 The logarithmic
scale option (see: \ref{option:logScale})  can be activated at the
analysis step of {\tt MC-TESTER}.
A nice illustration of this functionality is
the validation plot for {\tt PHOTOS} in the case of QED radiation from 
$W^+ \to \mu^+ \nu_\mu$ final state, see  fig.~\ref{logfalse}a.
The compared distributions\footnote{Obtained respectively from the versions of PHOTOS where
the matrix element is used (and where it is not).} 
 are visible only in the first few bins of the histogram.
If a logarithmic scale is used
(see fig.~\ref{logfalse}b),
one can see that the distribution extends over all kinematically allowed 
spectrum. 
Note that the ratio (thick black line) is still in linear scale marked on 
the left 
side of the plot. Calculation of Shape Difference Parameter is not changed
if logarithmic scale is used.

In  case of  tests where the coverage of some regions of  phase space
is  enhanced because of resonances,   distribution  properties
are best  visualized  by distributions of
Lorentz invariant masses constructed of all possible  sub-groups of final
 state momenta. This is the key concept of the {\tt MC-TESTER} methodology.

That was the case of $\tau$ lepton decays. Typically, 
one of the 
constructed invariant mass distributions 
 was peaked  around the position of the intermediate state resonance.
The tails of the distributions are better populated when invariant masses
are used directly, rather than 
 higher powers of these invariant masses.
To adopt for that type of applications, 
two  options are introduced into {\tt MC-TESTER}:
the ability to plot given powers of the inv. mass, (see: \ref{option:massPower}), 
and to scale the mass to its maximum possible value\footnote{The actually 
used power of the mass  appears in the plot name as {\tt Mass(2)} or  {\tt Mass(1)}
respectively if mass square or mass itself is histogrammed.} 
(see: \ref{option:massScaling}).

\begin{figure}[!ht]
\setlength{\unitlength}{0.1mm}
\begin{picture}(1600,800)
\put( 375,750){\makebox(0,0)[b]{\large }}
\put(1225,750){\makebox(0,0)[b]{\large }}
\put(5,100){\makebox(0,0)[lb]{\epsfig{file=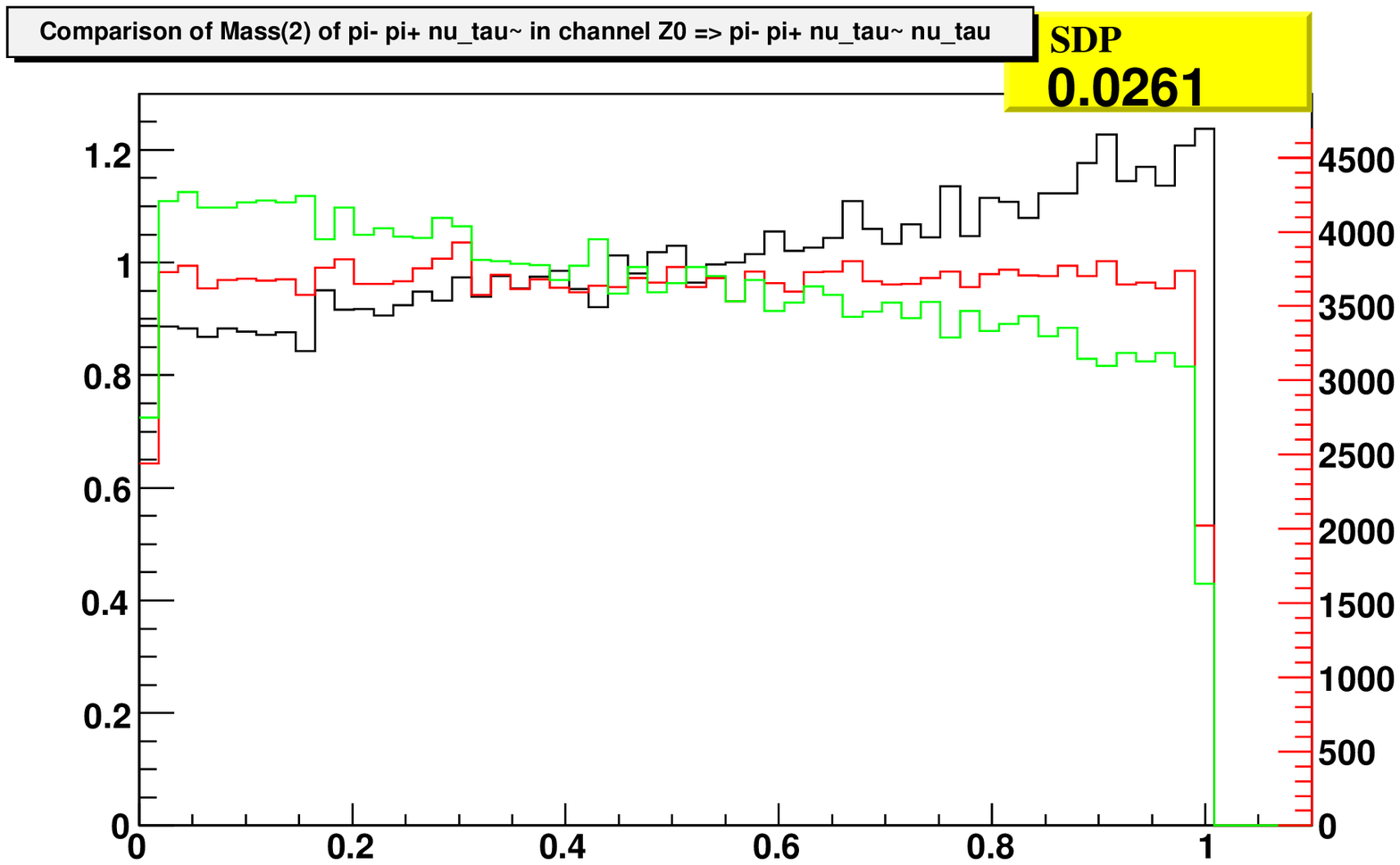,width=75mm,height=60mm}}}
\put( 55,75){\makebox(0,0)[lb]{a}}
\put(905,100){\makebox(0,0)[lb]{\epsfig{file=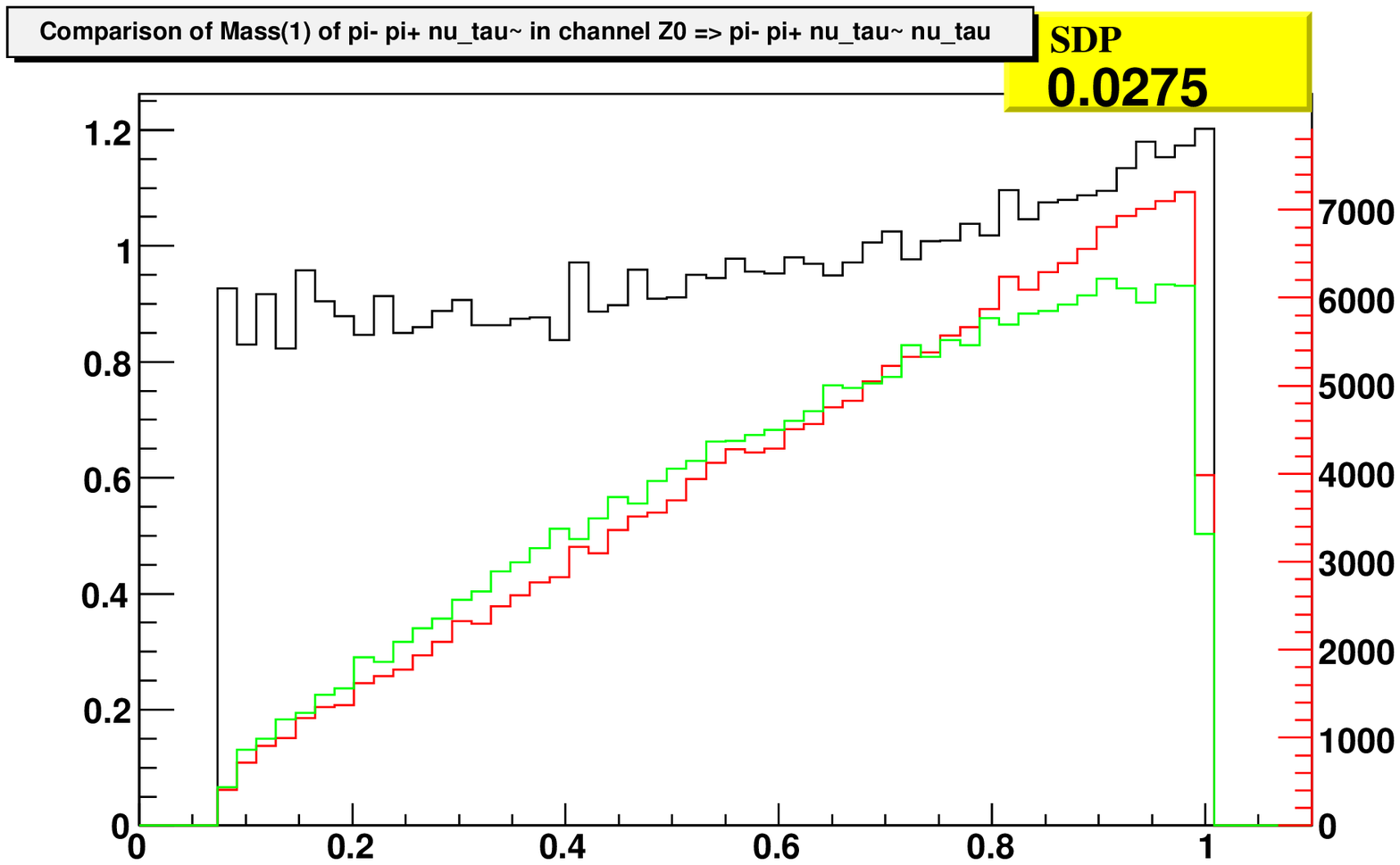,width=75mm,height=60mm}}}
\put(955,75){\makebox(0,0)[lb]{b}}
\end{picture}
\caption 
{\it Example of interesting benchmark for physics. In the decay
$Z \to \tau^+ \tau^- \to \pi^+ \pi^- \nu \bar \nu$, the $\pi^-$ energy spectrum in the Z rest-frame
has identical shape as the distribution of  the invariant mass squared 
 of $\pi^+ \pi^- \bar \nu$, see plot a. 
This linear distribution can be used to measure decaying $\tau^-$ polarization. 
The invariant mass is scaled to its kinematically allowed maximum. This
is convenient when the mass of the decaying object is not constant.
In plot b, the mass square option of\; {\tt MC-TESTER} 
parameters is not used. In principle such a test is equally good to validate spin 
effects, but any discrepancies are more difficult to understand.
 }
\label{massquare}
\end{figure}

The histograming of mass squared is useful  if one is interested in eg. spin
effects of $Z\to \tau^+ \tau^-, \tau^\pm \to \pi^\pm \nu $ decays.
The slope of the $\pi^-$ energy in the Z rest-frame is proportional to the
$\tau^-$ polarization, see fig.~\ref{massquare}a. This spectrum is identical
to the spectrum of invariant mass squared of the $\pi^+ \pi^-  \bar \nu$
system, and such distribution is now straightforward
 to study with {\tt MC-TESTER}.
For convenience we normalize the spectrum in proportion to its maximum
possible value, that is the invariant mass of the system of all ($Z$)
decay products.

Without using those new options, physically equivalent plots could still be
obtained, just plotting invariant mass, see fig.~\ref{massquare}b. It is however far more difficult
to interpret differences observed between compared Monte Carlo generators.

\newpage

\section{Package organization (update for ref.~\cite{Golonka:2002rz})}
\label{sec:package-organization}

This section contains technical details concerning {\tt MC-TESTER} and should be 
used (together with reference \cite{Golonka:2002rz}) as a quick guide book.
Further details may be found in the Appendix and in files placed 
in the {\tt doc/} subdirectory.
\subsection{Directory tree}
\label{dir-struct}

\begin{description}
\item [doc/] - contains documentation.
\item [examples-F77/] - includes example programs in {\tt F77}:
        \begin{description}
                \item [tauola/] - using the {\tt TAUOLA} generator;
                \item [pythia/] - using the {\tt PYTHIA/JetSet} generator;
                \item [pre-generated/] - results of generation with high statistics.
        \end{description}
\item [examples-C++/] - examples in {\tt C++}:
        \begin{description}
                \item [pythia/] - example for the {\tt PYTHIA 8.1} generator;
		\item [evtgenlhc/] - example for the {\tt EvtGenLHC} generator (CERN GENSER version).
	\end{description}
\item [analyze/] - analysis step is performed in this directory,
        the analysis code is contained in a set of \texttt{ROOT} macros.
        \begin{description}
                \item [prod1/] -  ({\tt mc-tester.root}) data file obtained from
                                 the generation phase  with the first generator  should be put here,
                \item [prod2/] -  ({\tt mc-tester.root}) data file from 
                                the results obtained with the second generator should be put here,
                \item [booklet/] - is created during the analysis step. It contains the result
                histograms in the form of {\it .eps} files.
        \end{description}
\item [HEPEvent/] - includes universal {\tt C++} interface to {\tt F77} event records
        ({\it i.e.} {\tt HEPEVT},{\tt LUJETS},{\tt PYJETS}).
\item [HepMCEvent/] - source code for {\tt libHepMCEvent} ({\tt MC-TESTER} 
		     interface libraries to the {\tt HepMC} event record);
\item  [include/]  - {\tt links} to {\tt C++} include files.
\item  [lib/] - contains compiled libraries needed by {\tt MC-TESTER}.
        Both the static and dynamic libraries are provided.
        
\item [src/]  - contains the source code for {\tt MC-TESTER}.
\item [platform/]    -  platform-dependent support files;
                        currently only for {\tt Linux}.
\end{description}

\subsection{Libraries }
The 
{\tt MC-TESTER} source code is build into three libraries:
{\tt libMCTester}, {\tt libHEPEvent} and {\tt libHepMCEvent}. These libraries may be found in 
the {\tt lib/} directory after installation.

The library {\tt libMCTester} contains all the code needed by the generation step; it is also required
at the analysis step, because it contains routines for the calculation of the {\tt Shape Difference 
Parameter}.

The {\tt libHEPEvent} library contains a unified interface for various 
{\tt F77 HEP} Monte Carlo
event record standards. 
In the current version of {\tt MC-TESTER}, it 
provides unified access to the {\tt HEPEVT}, {\tt LUJETS} and {\tt PYJETS} 
standards, enabling {\tt MC-TESTER} to be used with a variety of Monte Carlo 
event generators,
based on those
 event record standards. 

{\tt MC-TESTER} was recently extended to include the {\tt libHepMCEvent} library. This
library extends {\tt MC-TESTER}'s event record interface
defined in {\tt libHEPEvent} by adding support for the  {\tt HepMC} 
(versions greater than 2.0).


The source code of {\tt libMCTester} is placed in the {\tt src/} directory; {\tt libHEPEvent} is stored
in the {\tt HEPEvent/} directory; and {\tt libHepMCEvent} is stored in the {\tt HepMCEvent/}
directory.

\subsection{Format and syntax of the {\tt SETUP.C} file}
\label {SETUP.C}

The {\tt SETUP.C} file is a \texttt{C++ ROOT} macro file, which controls 
{\tt MC-TESTER}'s settings.
It is read and executed during initialization of both phases 
of a {\tt MC-TESTER} run: the generation and the analysis. Nothing, 
except points discussed in the Appendix, require further explanation here.
Documentation for the first version of the program is up to date.

\subsection{How to make {\tt MC-TESTER} run with other generators}
\label{other-generators}

\subsubsection{The case of {\tt C++} }
\label{other-gen:C++}
The infrastructure for connecting {\tt MC-TESTER} to a {\tt C++} generator was in place
in early versions of {\tt MC-TESTER}, but has recently been extended to allow analysis of 
C++ event records in {\tt HepMC} v2.0 and greater. As in the case for F77 code, the generator 
program should be linked to the {\tt MC-TESTER} libraries: {\tt libMCTester} and 
{\tt libHEPEvent}, as well as a subset of {\tt ROOT} libraries. Configuration of 
{\tt MC-TESTER} can be done either through a {\tt SETUP.C} file (see Section \ref{SETUP.C}) 
or directly in the body of the generation code (see Appendix \ref{appendix.C++}).


\begin{itemize} 
\item {\bf C++ generation program with {\tt FORTRAN} event record}\\
  All event record standards included in the current version of the {\tt HEPEvent}
  library, i.e. {\tt HEPEVT}, {\tt LUJETS} and {\tt PYJETS} may directly
  be used in a user's {\tt C++} code.  It is sufficient to issue calls to the
  following three functions
  inside the tested Monte Carlo analysis. 
  \begin{enumerate}
  \item {\tt MC\_Initialize()}: initializes {\tt MC-TESTER}. All changes
    to the {\tt Setup} should be commenced before a call to this function is invoked.
  \item {\tt MC\_Analyze()}: performs the analysis of
    the event record specified in the {\tt Setup::EVENT} variable; 
  \item {\tt MC\_Finalize()}: writes the results to the output file.
  \end{enumerate}
{\tt Generate.h} contains definitions for the functions listed above, and should be 
included in the generation program.
      
An example of running {\tt MC-TESTER} with {\tt EvtGenLHC} (C++ with {\tt HEPEVT} event 
record standard) can be found in the directory {\tt examples-C++/evtgenlhc/}.

\item{\bf C++ generator with {\tt HepMC} event record}\\
To analyze {\tt HepMC}\footnote{Only {\tt HepMC} version 2.0 or greater is supported.} event records
the additional interface library {\tt libHepMCEvent} should be linked to the generation code, and
{\tt HepMCEvent.H} as well as {\tt Generate.h} needs to be included.
Inside the main (and to be tested) event generation program, {\tt MC-TESTER} can be called in the following way:
\begin{enumerate}
    \item{\tt MC\_Initialize()}: initializes {\tt MC-TESTER}.
    \item{\tt HepMCEvent temp\_event(HepMC::GenEvent event)}: creates 
      an {\tt MC-TESTER} interface event to the {\tt HepMC::GenEvent} type event;
    \item{\tt MC\_Analyze(\&temp\_event)}: performs an analysis of the 
      {\tt HepMC::GenEvent} passed to temp\_event. The variable
      {\tt Setup::EVENT} in {\tt SETUP.C} will be ignored;
    \item{ \tt MC\_Finalize()}: writes the results to the output file.
\end{enumerate}

An example of {\tt MC-TESTER}'s application to {\tt HepMC} events can be found
for {\tt PYTHIA} 8.1 in the directory {\tt examples-C++/pythia/}.

Each Monte-Carlo generators could potentially have its own set of particle status 
codes to record the types of processes a particle can undergo. For example in
{\tt PYTHIA 8.1}, negative integers are used for decaying particles, while in 
{\tt HERWIG} positive integers are used. Therefore, it is important to document
the way in which {\tt MC-TESTER} interprets {\tt HepMC} status code information to conclude 
if a particle is stable, decayed, or history/documentation (consistent with the 
{\tt HEPEVT} standard codes of 1,2,3 respectively). The following definitions have 
worked successful with a variety of C++ MC Programs including {\tt PYTHIA 8.1} 
and {\tt Herwig++}. 

\begin{itemize}
  \item {\bf Stable:} If a particle has status code 1 or no end vertex of type 
    {\tt HepMC::GenVertex} it will be treated as stable by {\tt MC-TESTER}. 
    The particle will be placed in the list of daughters (see Section \ref{sec:Lists}.) 
    and the decay chain will not be traversed beyond this particle.
  \item {\bf History:} If a particle has status code 3 it will be treated as 
    documentation and {\tt MC-TESTER} will not place it in the list of daughters. 
    The decay chain will not be traversed beyond this particle.
  \item {\bf Decayed:} All other particles will be classified as decayed. The
    treatment of these particles by {\tt MC-TESTER} depends on the configuration
    of SETUP.C (see also  \ref{appendix.C++}). 
\end{itemize}

For the use of status codes beyond the above definitions, {\tt MC-TESTER} must be run 
with a modified version of the event record (see Section \ref{sec:HepMCaid}).

Filtering of the daughter list can be achieved using {\tt UserTreeAnalysis} (see 
Appendix \ref{option:user_event_analysis}).

\end{itemize}




\subsection{How to make {\tt MC-TESTER} run with on-flight modified event records}
\label{sec:HepMCaid}

Often we want to pass into {\tt MC-TESTER}, not the exact event tree as stored
in the event record, but a modified one. To date,
{\tt UserTreeAnalysis} (see \ref{option:user_event_analysis}) has been used
for this purpose. It provides a solution which works quite well, and hence will be 
discussed in the next section. Numerous options can be introduced in this way. 
However, methods are limited to manipulation during creation of 
the list of stable (endpoint) objects, originating from the analyzed object, or later. 
Some analysis of the object itself can be introduced as well.
The access to its origin or eg. kinematical properties may be interesting.


This requires that the event record logically matches our requirements. 
This is not necessarily always the case and one is tempted to perform some 
adjustment. In short, one may be interested in {\tt HepMC} to {\tt HepMC}
translation. 

We have realized that such modules can be of interest for other applications 
independent from {\tt MC-TESTER} as well. We will not discuss the related 
topic here at all. For practical use one need to call a method, which 
creates a temporary copy of the modified HepMC event accordingly.

\noindent
The sequence of calls, which can serve as a possible template for future work,
may look as follows:

{\tt HepMCaid HepMCnew; 

HepMCnew.SetOption(...);

.....

HepMCnew.ModifyEvent(HepMCold);

MC\_Analyze(HepMCnew);  }

\section{The use of lists in MC-TESTER}
\label{sec:Lists}

{\tt MC-TESTER} was developed over many years of experience in extracting data 
from different variants of event records. This section  addresses the readers who are
interested in this aspect of activity. Our past experience may be of future 
use,  it may  
explain some of the reasons behind our program design too.


Even though the idea of a standardized event record structures has been in place for a 
long time, technical problems  arose. 
For programs coded in {\tt FORTRAN}, event records (such as {\tt HEPEVT} or
{\tt LUJETS/PYJETS}) were typically expressed as integer-indexed arrays, with every
index representing a single item, such as a particle. To express relationships
between the particles or the ordering, one had to use the arrays of indices, as
{\tt FORTRAN} did not provide for more complex data structures such as lists. The
emulation of such structures through the arrays of indices used in the standards
such as {\tt HEPEVT} soon reached its limits: the arrays that were designed to store
the indexing information for doubly-linked lists representing the decay tree
started to fail, due to information overload: the place designated for the
storage of pointers in the emulated decay tree structure was re-used to store
additional information. 
The - initially easy - way to navigate through the
tree-like structure of the decay cascade became inconsistent,
generator-dependent and non-trivial to interpret. Moreover, to compensate for the
oversimplified model of the event structure, the authors of the event generators
started to introduce their own conventions, to fit additional information about
the intermediate, internal states of the particles and objects, which could not
be expressed directly\footnote{As a result,
 one could see in the encoded decay trees, cases
such as a $\tau$ particle decaying to another $\tau$ and a photon, and then this
second $\tau$ decaying to yet some other particles with both of the $\tau$'s being
actually two instances of the same particle, and the fact of having them listed
twice in the event record was to express their state at two distinctive stages
of event construction, and express the bremsstrahlung processes
of $\tau$ production rather than decay.}.

Energy-momentum conservation, at the tree branchings, was not always
assumed; branching points had to be `understood' as part of the larger group. The
original arrays that stored mother and daughter pointers to encode the
decay process started to serve multiple purposes, depending on the context:
either (as originally) to indicate the decay processes, or to encode some other
relationship. Needless to say that for the programs such as 
{\tt PHOTOS}~\cite{Barberio:1990ms,Barberio:1994qi,Golonka:2005pn,Golonka:2006tw}, which relies
on consistent information about the decay cascade structure to extract
sub-trees, or {\tt TAUOLA}~\cite{Pierzchala:2001gc,Golonka:2003xt}, 
which needs to extract the information about the hard
process, it has become extremely difficult to maintain a consistent generic
and reliable interface to other event generators. 

The same type of problems
obviously manifested in the large detector-simulation chains, where the input
from a theory-rich physics Monte Carlo event generator needed to be combined
with the phenomenological description of the detector processes, to give
observables such as energy-deposits and hits, which are in turn re-processed to
reconstruct the original process. Extracting the "signal" process from an event
record structure with overloaded data, full of generator-specific conventions
made this task, again, difficult and prone to errors; additional analysis steps
(which merely had to compensate for the insufficient data model of the event
record) had to be fitted to enable the use of each new event generator.

In the earliest version of {\tt MC-TESTER}~\cite{Golonka:2002rz} we suffered from the same type of problems
with index-based navigation through the event records. However, since version
1.1, we decided to take advantage of possibilities given by the C++ programming
language, and employ more complex data structures in the processing performed by
{\tt MC-TESTER}. The {\tt libHEPEvent }interface was extended with the abstraction of "particle
list". The particle list ({\tt HEPParticleList} class) could store pointers to any
number of objects representing the particles in the event record. A set of
methods, modelled on the concept of iterators from the C++ Standard Template
Library was added as well, to facilitate the navigation in the list object,
effectively replacing the native constructs of index-based "FOR" or "DO" loops.
The previously used methods to navigate through the event tree using the indices
were replaced with the list: each type of event record (supported by
{\tt libHEPEvent}) had a new method that returned a list of child particles,
and the associated iterator object was implemented in such a way that
all the generator-specific conventions were hidden in it, presenting
a clean and simple-to-use interface.  The code of {\tt MC-TESTER} was modified to
make use of these new constructs - it gained significantly in clarity and
stability: not only the previously-used, index-based syntax being
replaced with the constructs native to Object-Oriented languages, but also the
possible dependencies (or incompatibilities) could be delegated to be
served within another module ({\tt libHepEvent}), making the code much easier to
maintain. Ultimately, this also enabled the implementation of the {\tt HepMCEvent}
interface, and the use of {\tt HepMC}-based event generators with {\tt MC-TESTER},
in a straightforward way.

The mechanisms that extracts the lists of "daughter" particles, which is
currently implemented in the {\tt libHEPEvent}, still does not address the more
fundamental problem of over-simplification in the so-far proposed event record
structures, including {\tt HepMC}: the inability to express other types of
relations between particles necessary for  more complex models of processes
where for example quantum interference need to be included.
 Up to now,
there is only one type of "relations", being mother-daughter relations in the
old event records, or "interaction vertices" or "blobs" in the newer ones: they
only express the relation in the decay cascade. Other types of relations or
processes still need to be "emulated" by employing special conventions, or
additional, often non-physical objects (many instances of the same particle,
etc). As already discussed, the resulting data structure is difficult to
interpret:  objects as physical entities,  the actual processes taking place during (often
multi-step) event generation, and  to implement "content
enriching" generators, such as  {\tt TAUOLA} and  {\tt PHOTOS}, which add to an already
(partially) generated event, stored in the event record. Up to this point, we
treat the exercises with list-based methods for "reinterpretation" of the
event-record data, as an initial seed for a more concentrated effort to provide such
"re-interpreting" code in the near future, targeted in particular as
helper utilities for the software of large experimental collaborations. Additional
ingredients for such utilities are provided by the experience we gained with
the "user tree analysis" feature of  {\tt MC-TESTER}, documented in Appendix \ref{option:user_event_analysis} and Section \ref{sec:HepMCaid}, where we
extract/re-construct/correct a fragment of the event record "on the flight", and
present it as the input to  {\tt MC-TESTER}, rather than using the original event
record. Such an approach, with re-engineered, re-interpreted event data, created
on-the-flight, using a simple-to-use, pluggable script/macro files has an
additional advantage: the original event records remains unmodified, and could
still be accessed.

{\tt MC-TESTER} is not the only of our projects, where the problems 
discussed above have to be addressed.  {\tt TAUOLA universal 
interface} \cite{Golonka:2003xt} and  {\tt PHOTOS} \cite{Barberio:1994qi} 
represent further examples. Our program is devoted to tests, this is why 
it was worked out before the other two.

Similar solutions to problems as those discussed here \cite{Was:2001te}, 
are possible.
It seems that in this respect, the case of {\tt TAUOLA universal interface} 
is easier than {\tt PHOTOS}.
New objects need to be added to the end points of the otherwise unmodified 
tree. The prototype solution, based on {\tt HepMC} exist already 
 \cite{Davidson:2010rw}.

The standard concept of C++ lists 
can be used for minor practical adaptation in {\tt MC-TESTER}  too. 
For example, at the time of list  creation  one can force some particles 
to be treated as stable, and its consecutive daughters ignored.

\section{Example of advanced {\tt MC-TESTER} use: \\
 benchmarks for spin correlations in heavy object decays. }

\label{sec:Benchmarks}
One may have the impression that the modifications introduced into the present 
release of the package are minor and consist of simple improvements in 
the graphical representation of the output and purely technical reorganization 
thanks to the use of C++ lists. 

To some degree this is true, but  other  changes were introduced 
because of pressure from applications. In the present chapter, let us show, 
how program modifications can be used for non-trivial practical applications.

It is quite common that information stored in the event record is too large. 
For example individual soft photons which remain undetectable are present. Not only
they do not influence  the detector response at all, but they exhibit technical 
aspects of eg. infrared regulators of QED bremsstrahlung. In response, 
{\tt MC-TESTER}
should ignore (or group together with other particles), 
those photons while analyzing decays. Otherwise comparisons 
of different Monte Carlo programs would be dominated by the technical aspect of 
the implementation of infrared regulator; {\tt MC-TESTER} operation need to be 
adopted for this, see eg. \cite{Golonka:2005pn,Golonka:2006tw}. 

Another example where the event tree may need to be simplified for validation is 
if spin correlations are appropriately introduced into various production 
processes.  Let us use as an example\footnote{Spin correlations  in 
  decays of $W,H,H^\pm$ into $\tau$ lepton(s)  are nearly identical}
 $ pp \to Z/\gamma* + X$, 
$Z \to \tau^+ \tau^-$. It is convenient to start the test by restricting 
$\tau$ decays to the simplest decay mode, that is $\tau^\pm \to \pi^\pm \nu$,
and look at distributions in combined decay $Z \to \pi^+ \pi^- \nu \bar \nu$.
In this case the effects of spin correlations are largest. 
The distribution of the $\pi^-$ energy spectrum (in the $Z$ rest-frame),
manifests the $\tau$ polarization through its slope (see 
fig  \ref{massquare}a). 
Fortunately, this frame dependent spectrum is equal to 
the  distribution of the  invariant mass squared of $\pi^- \pi^+ \bar \nu$.
This distribution can be obtained in {\tt MC-TESTER} thanks to the new options of 
histograming invariants squared~(\ref{option:massPower})   and 
automatic scaling of the histogram range to the kinematically allowed  
maximum~(\ref{option:massScaling}).

Final state activities will lead to  $Z$ decays
where the $\tau$-pair is accompanied eg. by bremsstrahlung photons or soft hadrons. 
One may want to ignore this soft radiation in the test, 
or quite  contrary -- look only at these cases, to verify if soft emissions did
 not result from configurations of faulty spin correlations.
Finally one may want to check decays of  $Z$ of high $p_T$ only. 

On the other hand all such variants of non-standard {\tt MC-TESTER} analysis were rather easy  to include 
into our example of the {\tt UserTreeAnalysis} (see.~\ref{option:user_event_analysis}),
but we assume that in the future other options may also become useful. 
On the other hand some interesting variants of the {\tt UserTreeAnalysis} method 
may not be possible using decay product lists alone. 
For example if one would be interested in decays
of $Z$s originating from an object $X$ or accompanied in $X$ decay by eg. another
$Z$ or  top quark. For that purpose some other methods following the idea of
{\tt userEventAnalysis} \cite{Golonka:2002rz} or HepMCaid (see~\ref{sec:HepMCaid})
may  be useful.

\subsection{Default UserTreeAnalysis }
\label{sec:UserTreeAnalysis}

 {\tt UserTreeAnalysis}, is included in the source code of MC-TESTER
thus can be loaded with the library libMCTester. The user can create his 
own verson of the method named eg. 
{\tt MyUserTree}\footnote{See Appendix \ref{option:user_event_analysis} for details.}, 
and load it  as a pre-compiled C++ macro instead. The parameters of  
the built-in {\tt UserTreeAnalysis}  named   {\tt "UserTreeAnalysis"},
have the following meaning:

\begin{enumerate}
\item \texttt{params[0]=0.05} minimum value of the variable used to discriminate
          soft particles as a fraction of the decaying particle mass.
\item \texttt{params[1]=0} maximum number of possible soft particles retained, 
   even if passing a threshold of the previous option. 
\item \texttt{params[2]=0} type of variable used in discrimination 
   \begin{enumerate}
     \item  0 - energy in the decaying particle rest frame
     \item  1 - energy in lab frame
     \item  2 - $p_T$ in lab frame
   \end{enumerate}
\item \texttt{params[3]=0}
   \begin{enumerate}
     \item  0 - removed particles are simply ignored
     \item  1 - removed particle momenta are added to the momenta of charged
            ones. For details see \cite{Golonka:2002rz}.
   \end{enumerate}
\
\item \texttt{params[4]=22}  PDG Id's of the particles to be removed. 
Repetition of this  parameter is allowed for \texttt{params[5]}, \texttt{params[6]}  etc. 
\end{enumerate}

If parameters are not
initialized, the default values, as given in the points above, are used. For the example, our
 method defined three histograms for the properties of the decaying particle: its 
$p_T$, pseudorapidity and  azimuthal angle $\phi$. They are included, during the analysis step, 
in the ``User Histogram'' section of the {\tt  MC-TESTER} booklet.

This simple method summarizes and extends the technical aspects of tests we have developed in  papers \cite{Golonka:2005pn,Golonka:2006tw}.

\section{Outlook}
\label{sec:outlook}
We have demonstrated that   {\tt MC-TESTER} may be useful for tests of 
 libraries of particles decays, as well as for tests of their interfaces
(see fig.~\ref{massquare} and refs.~\cite{Pierzchala:2001gc,Davidson:2010rw} for example).

The updated version of the package was found \cite{Golonka:2005pn,Golonka:2006tw} to handle
 well cases where physically spurious information (eg. on soft photons)
need to be ignored. This avoided unphysical discrepancies between results from different 
programs. Moreover,  adaptations of the program may lead to 
a new spectrum of applications, which, as discussed in Sections \ref{sec:HepMCaid}, \ref{sec:Lists} and  \ref{sec:Benchmarks}, may find applications independent of the future evolution of the {\tt MC-TESTER} software project.

Even with the present enrichment of functionality, the tests performed by {\tt MC-TESTER}
are not complete from the physics point of view. The program has also some technical limitations.
In the following we list these points, which may be addressed in future
 versions of {\tt MC-TESTER} and require stating.

\begin{enumerate}

\item
The program does not analyze distributions in Lorentz invariants built with the 
help of the totally antisymmetric (Levi-Civita) tensor. It is thus blind to some 
effects of parity non-conservation. 
\item
Information on the spin state of the decaying particle is usually not 
available in event record structures such as {\tt HepMC}.
To keep  {\tt MC-TESTER}  modular, and to avoid 
a multitude of options, we ignore
effects of decaying particle polarization. 
\item
The main advantage of    
{\tt MC-TESTER} is that it can be used with `any' production generator in 
an automated way, providing a tool for quick tests. However, the final state 
event record has to be stored in one of the following structures:
common blocks {\tt HEPEVT, LUJETS, PYJETS} \cite{PDG:1998,Pythia} of {\tt FORTRAN} or
{\tt HepMC} objects of {\tt C++}.
\item
If multiplicity of the particular decay channel is very high and/or
there is a lot of decay channels, the program may
find it difficult to allocate memory. 
An analysis of a decay channel with 8 or more decay products results in
thousands of histograms, which causes output files to be large and the analysis
step to be long.
%
Hard coded limits have been implemented:
Histograms will only be created for the first 200 decay channels found and only if the 
multiplicity of decay products is smaller than 8.

\item Some of {\tt MC-TESTER}'s options, especially {\tt MyUserTree} method,
see Section \ref{sec:UserTreeAnalysis},
may be difficult to use within large systems like Athena\footnote{http://atlas.physics.utoronto.ca/Members/bguo/setup-athena-12-0-0-at-cern-machines} of the ATLAS 
collaboration. This point will also need to be investigated after the release 
of the present version of our program.  It requires interaction with the users.

\end{enumerate} 

 The main purpose of {\tt MC-TESTER} is to analyze sub-trees
starting from the objects of a given {\tt PDG} identifier without any concern
of its origin, and without pre-selecting the type of distributions created.
This is  why there is  
complementarity between our approach and the one of {\tt Rivet} \cite{Rivet}.
 The latter is
designed to produce 
simulated distributions which can be directly compared to measured data for 
validation and tuning purposes.
 The {\tt MC-TESTER} strategy is to
 test decays on a technical level at the event record content first, 
rather than to start from already pre-identified quantities of physics interest.
Only later one may, but with constrained possibilities only,
 turn to physically interesting quantities. 

Updates introduced to the program after  version  1.23 become public are described in 
Appendix \ref{appendix.B}.

\section*{Acknowledgments}
We thank Elzbieta Richter-Was, Alberto Ribon,  Judith Krantz and 
Zhonghua Qin for comments 
on the program organization and documentation. Nadia Davidson would like to thank the
``Marie Curie Programme'' for her fellowship. Partial support of 
Polish-French collaboration no. 06-124 
 within IN2P3 through LAPP Annecy during final completion  
 of this work is also acknowledged.

\appendix
\section{Appendix: {\tt MC-TESTER} setup and input parameters (update for ref.~\cite{Golonka:2002rz})}
\label{appendix.A}

The values of the parameters used by {\tt MC-TESTER} are controlled using 
the {\tt SETUP.C} file. Some parameters may also
be controlled using {\tt FORTRAN77} interface routines 
 or C++ methods (Section \ref{appendix.C++}).
This provides runtime control over all parameters, yet allowing the user
not to have {\tt SETUP.C} at all.
One should note that {\tt SETUP.C} always has precedence over the default values set 
using {\tt F77} or C++ code: it is always looked for in the execution directory.

\subsection{Definition of parameters in the {\tt SETUP.C} file}
\label{appendix.Setup-parameters}
There are three sets of settings inside {\tt MC-TESTER} to be distinguished: 
the ones
specific to the generation phase, the ones specific to the analysis phase and the ones
that are used in both phases\footnote{Some parameters from the generation 
 phase (i.e. the description of generators) are stored inside
 an output data file. However, again for reasons of runtime control, their 
 values may be altered at the analysis time using the {\tt SETUP.C} file in 
 the analysis directory.}. We describe only new features, quoting the scope of their use.

\subsubsection{Setup::UserTreeAnalysis}
\label{option:user_event_analysis}
Type:~char* 

Scope:~generation 

Default: null

DESCRIPTION: The name of a function that allows modification of the list of stable particles before
histograms for the decay of {\tt MC-TESTER} analyzed object are 
defined/filled in.

IMPORTANT: The name that is attributed (eg. "MyUserTree") must be a
valid method name existing in a C++ script file located in the
working directory. The script must be given the same name as the method,
and ended with a ".C" suffix. For example, for the "MyUserTree" method,
the script filename would be {\tt MyUserTree.C}.
For further information on running and compiling scripts on the fly see
{\tt README.UserTreeAnalysis} in the {\tt MC-TESTER/doc/} directory.

Example of use:

\texttt{ Setup::UserTreeAnalysis = "MyUserTree";}

or for the version compiled and present in {\tt libMCTester} library:

\texttt{ Setup::UserTreeAnalysis = "UserTreeAnalysis";}

In this case, the UserTreeAnalysis.C is not needed, as the
built-in UserTreeAnalysis routine will be used.

Parameters can be passed to the function. For example

\texttt{Setup::UTA\_{}params[0]=0.05; }

\texttt{Setup::UTA\_{}params[1]=0; }

\texttt{Setup::UTA\_{}params[2]=0; }

\texttt{Setup::UTA\_{}params[3]=0;  }

\texttt{Setup::UTA\_{}params[4]=22; }

\texttt{Setup::UTA\_{}params[5]=111; }

\texttt{Setup::UTA\_{}nparams=6;}

\noindent
will pass {\tt nparams=6} parameters to the function.
For the actual meaning of the parameters if passed into
{\tt UserTreeAnalysis} as present in the library,  see section \ref{sec:Benchmarks}.

\subsubsection{Setup::mass\_power }
\label{option:massPower}

Type:~ int 

Scope:~ generation 

Default: 1

DESCRIPTION: This option changes the variable passed
for histograming, from invariant mass to a power of invariant mass, at the generation step.
It also modifies the title displayed on histograms from {\tt Mass(1)}
to an appropriate {\tt Mass(value)}, showing that the power of the mass has
been changed.

NOTE: Acceptable values: from 1 to 9. Due to properties of the Lorentz group 
when this option has value=2 it is particularly suitable for tests of spin polarization, 
see section~\ref{sec:UserTreeAnalysis}.

Example of use: 

\texttt{ Setup::mass\_power=2; //set histograms to invariant mass squared}

\subsubsection{Setup::mass\_scale\_on }
\label{option:massScaling}

Type:~ bool

Scope:~ generation 

Default: false

DESCRIPTION: This option scales invariant masses for all plots of the decay channel to invariant mass
constructed from all daughters combined. It scales the X values to the range (0,1). 

NOTE: When using this option consider setting default maximum bin value to 1.1,
for nicer graphical representation.

Example of use: 

\texttt{ Setup::mass\_scale\_on=true; //enables scaling of X axis }

\subsubsection{Setup::use\_log\_y }
\label{option:logScale}

Type:~ bool

Scope:~ analysis

Default: false

DESCRIPTION: Enables the use of logarithmic scale in all histograms
plotted by {\tt MC-TESTER}. Turning this option on will draw the histograms
 in logarithmic scale, and mark a logarithmic scale along the
right-hand-side Y axis. This option does not affect {\tt SDP} calculation or
the plot of the ratio of histograms, which remains linear. Its corresponding
linear scale is marked on the left-hand Y axis.

NOTE: This option, combined with previously presented defaults for
      UserTreeAnalysis can be particularly useful if infrared regulator
      sensitive particles, such as soft photons are present in the event records.
      See \cite{Photos_tests}.

Example of use:

\texttt{ Setup::use\_log\_y=true; //enables logarithmic scale on Y axis }

\subsubsection{Setup::rebin\_factor }
\label{option:rebinfactor}

Type:~ int 

Scope:~ analysis 

Default: 1

DESCRIPTION: One may want to define a large number of bins for the generation
scope of {\tt MC-TESTER}. The number of bins on the actual plots,
can be adjusted at the analysis step. The contents of consecutive 
"rebin\_factor'' bins are summed together. Calculation of the SDP parameter
is appropriately adjusted. 

NOTE: "rebin\_factor'' must be the natural divider of the number of bins 
declared during the generation scope of {\tt MC-TESTER}.

Example of use: 

\texttt{ Setup::rebin\_factor=3; //reduces no. of bins in all histograms by factor of 3}

\subsection{ {\tt C++} configuration of {\tt MC-TESTER}}
\label{appendix.C++}
The configuration of {\tt MC-TESTER} can be done directly in the main method
of the C++ generation program, without the need for a {\tt SETUP.C} file. This can be accomplished
by including the header file {\tt Setup.H} and setting parameters 
using the same syntax as described for {\tt SETUP.C} files 
(see original documentation ~\cite{Golonka:2002rz}).
Setup should be done before calling the function {\tt MC\_Initialize()}. 
Note that if parameters are set in both the generation
program and a {\tt SETUP.C} file, the values present in {\tt SETUP.C} will be given precedence.

\section{Appendix: updates to versions 1.24.2 and 1.24.3}
\label{appendix.B}
\subsection{Changes introduced in version 1.24.2}

       To address the problems that are typically faced when {\tt MC-TESTER} is
installed in a new environment, or a new platform, an automated configuration
step has been implemented in version 1.24.2. The configuration files required
to set-up/compile/run {\tt MC-TESTER} may be generated through a dedicated
configuration script, which facilitates the GNU autoconf \cite{autoconf}.

To set up {\tt MC-TESTER} using the new auto-configuration facility, proceed with
the following steps:

\begin{itemize}
  \item Execute {\tt ./configure} with additional command line options: \\
        {\tt --with-HepMC=<path>} provides the path to {\tt HepMC} installation directory 
(alternatively  {\tt HEPMCLOCATION} system variable has to be set). \\
{\tt --with-root=<path>} Path to {\tt root} binaries. \\
{\tt --with-Pythia8=<path>} Path to {\tt Pythia} version  8.1 or later (this generator is used by examples only)\\
        {\tt --prefix=<path>} provides the installation path. If this option is used {\tt include/} and {\tt lib/} directories will 
be copied to this {\tt prefix <path>} when {\tt make install} will be executed. If {\tt --prefix=<path>}
is not provided, 
the default installation directory   {\tt /usr/local} will be used
  \item Execute {\tt make}; this will build {\tt MC-TESTER}.
  \item To install {\tt MC-TESTER} into the directory specified at step 1) through the 
{\tt --prefix}
parameter, execute {\tt make install}; this will copy the include files
and libraries into {\tt include/} and {\tt lib/} sub-directories.
\item It is worth to mention that  {\tt ./configure} scripts only prepare  {\tt make.inc} files.  These files are rather short  
and can be 
easily modified or created  by hand: one can also rename {\tt README-NO-CONFIG.txt} 
to {\tt make.inc} and modify it accordingly to instructions provided inside 
the file.
\end{itemize}

Further changes and bug-fixes were implemented too. However they  do not require any changes in the way the program is used. Let us nonetheless list them here:

\begin{itemize}
  \item A bug resulting in faulty functioning of the script  {\tt ANALYZE.C} was fixed. Previously, when comparing decay samples which differed by several distinct
 channels, the program was occasionally crashing.
  \item A bug resulting in faulty functioning of {\tt UserTreeAnalysis} scripts was fixed. 
The program was crashing if
 {\tt MC4Vector} was used inside the script.
  \item All offending statements resulting in compilation errors if '-ansi -pedantic'  flags were activated have been removed now.
\end{itemize}

\subsection{  LCG configuration scripts; available from version 1.24.2  }

For our project still another configuration/automake system was prepared  by Dmitri Konstantinov and Oleg Zenin;
 members of the
LCG project \cite{LCG}. 

For the purpose of activation of this set of autotools-based installation scripts
enter  {\tt platform} directory and execute there {\tt use-LCG-config.sh} script.
Then, installation procedure and the names of the configuration script parameters will differ from the one 
described in our paper. Instruction given in  './INSTALL' readme file created by {\tt use-LCG-config.sh} script
should be followed. One can also execute {\tt ./configure --help}, it will 
list all options available for the configuration script.

A short information on these scripts can be found in {\tt README} of main directory as well.

\subsection{ Merging {\tt MC-TESTER} output files; available from version 1.24.3.}

 Interest in using the program on distributed systems, such
as the grid has been expressed on several occasions. 
This calls for new functionality: to
merge several {\tt mc-tester.root} files into a single one, 
corresponding to all event samples combined into one.

\noindent
The {\tt analyze/MERGE.C} script can be used for this purpose:
\begin{itemize}
  \item Enter the {\tt analyze/} directory.
  \item Execute {\tt root -b MERGE.C } (or {\tt root -b} and {\tt .L MERGE.C}).
  \item Type {\tt merge(<output file> , <input directory>/<first file> , [<pattern>])} (the last parameter is optional).
  \item Copy {\tt <output file>} into {\tt analyze/prod1/mc-tester.root} \\
        (or {\tt analyze/prod2/mc-tester.root}).
\end{itemize}
Example:\\
\\
{\tt root -b}\\
{\tt root [0] .L <path\_to\_script>/MERGE.C}\\
{\tt root [1] merge("out.root","samples/first.root","*.root")}\\
\\

The input to the script may consist of just the {\tt <input directory>} path 
where reside the {\tt .root} files to be merged. 
 Alternatively, the name of the first {\tt MC-TESTER .root} file to be merged ( {\tt <input directory>/<first file> }) 
can be explicitly given. Generator information will be taken from this first file.
The script will search the {\tt <input directory>} to merge all files matching 
the pattern. If no pattern is provided, the default pattern is {\tt mc-tester\_$*$.root}.
The {\tt <output file>} will  feature all histograms for decay channels including user defined histograms. 
Histograms for decay channels 
of the same 
name, found in different files will be summed together. 
The histogram bin count and axis range of the first occurence will be used. 

If histograms are found with a distinct axis range or number
of bins, compared to other histograms with the same name, then the
content of these files is ignored. However, if by mistake, the
particular input file contains data for tests of another particle's
decays, then all data from this file will be taken. All decay channels
for all particles under consideration will be listed in the .pdf file
constructed by {\tt MC-TESTER} at the analysis step. Information
printed on the front page might then be inconsistent, for example,
the overall number of entries or the overall number of channels will represent decays of all particles.
 If the interest will be expressed in future, an analysis step can be adopted to handle such cases with better front page of the booklet.

If the script is used outside the {\tt MC-TESTER/analyze/} directory, 
the {\tt MCTESTERLOCATION} system variable needs to be set to the {\tt MC-TESTER} root directory.
In addition, user can create and adopt his own copy of {\tt MERGE.C}  and use it instead of the default one. 
Note that our script cannot be used with a version of {\tt MC-TESTER} older than 1.24.3.

\bibliographystyle{utphys_spires}
\bibliography{tester}


\end{document}

\providecommand{\href}[2]{#2}

\end{document}